\documentclass[preprint,aps ,nofootinbib]{revtex4}
\usepackage[colorlinks=true,linkcolor=blue,urlcolor=blue,filecolor=black,citecolor=red,pdfstartview=FitV,pdftitle={},pdfsubject={},pdfkeywords={},pdfpagemode=None,bookmarksopen=true]{hyperref}
\usepackage{graphicx}
\usepackage{epstopdf}
\usepackage{amsmath}
\usepackage{amsfonts}
\usepackage{amssymb}
\usepackage{bm}
\usepackage{color}%
\usepackage{dcolumn}
\usepackage{slashed}
\usepackage{amssymb,ulem}
\usepackage{pdflscape}	
\usepackage{lineno}
\usepackage{slashed}

\newcommand{\f}{\begin{equation}}
\newcommand{\ff}{\end{equation}}
\newcommand{\fa}{\begin{eqnarray}}
\newcommand{\ffa}{\end{eqnarray}}

\begin{document}
\title{Chaotic dynamics of string around the conformal black hole}
\author{Da-Zhu Ma $^{1,2}$}
\thanks{mdzhbmy@126.com}
\author{Fang Xia $^{2}$}
\thanks{xf@pmo.ac.cn}
\author{Dan Zhang $^{3}$}
\thanks{danzhanglnk@163.com}
\author{Guo-Yang Fu $^{4}$}
\thanks{FuguoyangEDU@163.com}
\author{Jian-Pin Wu$^{4}$}
\thanks{jianpinwu@yzu.edu.cn}
\affiliation{
$^{1}$ School of Information and Engineering, Hubei Minzu University, Enshi 445000, China\ \\
$^{2}$ Purple Mountain Observatory, Chinese Academy of Sciences, Nanjing 210023, China\ \\
$^{3}$ Key Laboratory of Low Dimensional Quantum Structures and Quantum Control of Ministry of Education,
Synergetic Innovation Center for Quantum Effects and Applications,
and Department of Physics, Hunan Normal University, Changsha, Hunan 410081, China\ \\
$^{4}$ Center for Gravitation and Cosmology, College of Physical Science and Technology, Yangzhou University, Yangzhou 225009, China
}

\begin{abstract}

In this paper, we make a systematical and in-depth study on the chaotic dynamics of the string around the conformal black hole. Depending on the characteristic parameter of the conformal black hole and the initial position of the string, there are three kinds of dynamical behaviors: ordered, chaotic and being captured, chaotic but not being captured. A particular interesting observation is that there is a sharp transition in chaotic dynamics when the black hole horizon disappears, which is indepent of the initial position of the string. It provides a possible way to probe the horizon structure of the massive body.
We also examine the generalized MSS (Maldacena, Shenker and Stanford) inequality, which is proposed in holographic dual field theory, and find that the generalized MSS inequality holds even in the asymptotically flat black hole background. Especially, as the initial position of the string approaches the black hole horizon, the Lyapunov exponent also approaches the upper bound of the generalized MSS inequality.

\end{abstract}
\maketitle

\section{Introduction}

Because of the inherent non-linearity of General Relativity (GR), the chaotic dynamics has become one of the central attention in relativistic systems, where the chaotic phenomena have been deeply explored. The simplest dynamics is the geodesic motion of a test particle around a prescribed background, which is integrable in the generic Kerr-Newman background \cite{Carter:1968}. However, the chaos emerges in some complicated dynamical systems, for example the test particle around the Majumdar-Papapetrou geometry \cite{Dettmann:1994dj,Hanan:2006uf}, or particles near a black hole in a Melvin magnetic
universe \cite{Karas:1992}, or in a perturbed Schwarzschild spacetime \cite{Bombelli:1992,Aguirregabiria:1996vq,Sota:1995ms},
or in the accelerating and rotating black holes spacetime \cite{Chen:2016tmr}.
In addition, the motion of the particle in the certain potential has also been shown to exhibit chaotic phenomena, for example, the motion of charged particles in a magnetic field
interacting with gravitational waves \cite{Varvoglis:1992}.

On the other hand, we are interested in the dynamical system of the string.
Because of the inherent extended nature of the string, the motion of the string exhibits a more complex behavior. Even on the radially symmetric background, the dynamics of the string is also chaotic \cite{Frolov:1999pj,Zayas:2010fs,Ma:2014aha,Bai:2014wpa,Basu:2016zkr,Ishii:2016rlk,Cubrovic:2019qee,Ma:2019ewq}
. There is potential significance to explore the cosmic string around the astrophysical black hole \cite{Vilenkin:1994}. As pointed out in \cite{Vilenkin:1994}, up to the leading order thickness approximation, the cosmic string can be depicted by the Nambu-Goto action. The authors in \cite{Frolov:1999pj} study the dynamics of the circular cosmic string in asymptotically flat Schwarzschild black hole. Especially, they discuss the transition from order to chaos.

Another motivation to study string dynamics arises from the AdS/CFT (Anti-de Sitter/Conformal Field theory) correspondence \cite{Maldacena:1997re,Gubser:1998bc,Witten:1998qj,Aharony:1999ti}.
By studying the chaotic dynamics of the ring string around AdS-Schwarzschild (AdS-SS) black hole, the authors in \cite{Zayas:2010fs} propose that the positive largest Lyapunov exponent on the gravity side sets an appropriate bound for the time scale of Poincare recurrences on the gauge theory side.
Along this direction, the chaotic phenomena arising from the string motion around more general AdS geometries are revealed \cite{Ma:2014aha,Bai:2014wpa,Basu:2016zkr,Ishii:2016rlk,Cubrovic:2019qee,Ma:2019ewq}.

In this paper, we shall study the chaotic dynamics of the string in the Weyl conformal gravity.
The Weyl conformal gravity is a fourth-order gravity \cite{Weyl:1918pdp,MFathi:2019jgd}, which is a possible alternative to the standard second-order Einstein theory. The conformal Weyl gravity is invariant under a conformal transformation of the metric tensor
\begin{align}
\label{metric}
g_{\mu\nu}=\Omega^{2}g_{\mu\nu}\,,
\end{align}
where $\Omega$ is a function of the spacetime point. The Weyl conformal gravity can resolve the problem of flat galactic rotation curves \cite{Mannheim:1988dj}. Further, the Weyl conformal gravity can be used as an alternative to dark matter and explain the dark energy related phenomena \cite{Mannheim:2006,Nesbet:2013}. In addition, there are lots of the generalized studies of the Weyl conformal gravity, see for example \cite{Bambi:2016yne,Kasikci:2018mtg,Fathi:2019jgd,Takizawa:2020dja,Li:2020wvn,Fathi:2020sey,Fathi:2020sfw,Abbas:2020pzc,Konoplya:2020fwg}. Here, we shall study the string dynamics in the Weyl conformal gravity and explore the corresponding chaotic phenomena.

We organize the paper as what follows. In Section \ref{basis}, we present a brief review on the Weyl conformal gravity and the black hole solution.
And then, we work out the dynamical system of the string around the conformal black hole in Section \ref{sec-string}.
In Section \ref{sec-chaodyn}, we numerically solve the dynamical system and explore the properties of the string dynamics by chaos indicators.
Also we examine the generalized MSS (Maldacena, Shenker and Stanford) inequality, which is proposed in holographic dual field theory. The conclusions and discussions are presented in Section \ref{sec-con}.

\section{The conformal black hole}\label{basis}

We start with the following action
\begin{align}
\label{action}
S=\int d^{4}x\sqrt{-g}C_{abcd}C^{abcd}\,,
\end{align}
where $C_{abcd}$ is the conformal Weyl tensor. This theory is a fourth-order gravity theory. It is invariant under a conformal transformation as have been pointed out in the introduction. A static and spherically symmetric vacuum solution from the action \eqref{action} is given \cite{Mannheim:1991}
\fa
&&
\label{metric}
ds^2=-B(r)dt^2+\frac{dr^2}{B(r)}+r^{2}(d\theta^{2}+{\sin^{2}{\theta}}d\phi^{2})\,,
\
\\
&&
B(r)=1-\frac{\beta(2-3\beta\gamma)}{r}-3\beta\gamma+\gamma r-kr^{2}\,.
\label{metric-Br}
\ffa
There are three integral constants, $\gamma$, $\beta$ and $k$ in this solution.
When $\gamma = k = 0$, the above solution reduces to the Schwarzschild solution for a spherically symmetric source of mass $\beta=M_{0}$. The last term, i.e., the $kr^{2}$ term, plays the role of the effective cosmological constant and becomes important at cosmological distances. $\gamma$ is the characteristic parameter of the conformal black hole, also dubbed as MK (Mannheim and Kazanas) parameter \cite{Mannheim:1991}. When $\gamma=0$, this solution reduces to the Schwarzschild (Anti-)de Sitter solution. This theory allows one to describe flat rotation of galaxies without introducing the dark matte, for which $\gamma$ is of the order of the inverse of the Hubble radius.

Depending on the parameters, the conformal black hole exhibits rich structure.
\begin{itemize}
  \item When $\beta=0$, the black hole solution \eqref{metric} is conformally flat \cite{Mannheim:1988dj}.
  \item When $\beta\neq 0$, the conformal flatness of the background is broken. Therefore, the black hole solution \eqref{metric} can be seen as a massive body embedded in a conformally flat space \cite{Mannheim:1988dj}.
  \item The Newtonian term $1/r$ plays an important role when $r$ is small and it vanishes as $r\rightarrow\infty$, for which the other terms dominate \cite{Mannheim:1988dj}.
\end{itemize}

In addition, the conformal black hole also has rich horizon structures: two horizons, one horizon and no horizon depending on the parameters (see for example Refs.\cite{Abbas:2020pzc,Villanueva:2013gga,Turner:2020gxo} for detailed discussions). The Hawking temperature of the conformal black hole is given by 
\begin{align}
	T=\frac{-2 k r_h^3+2\beta+\gamma r_h^2-3\gamma\beta^2}{4\pi r_h^2}\,.
	\label{Tem}
\end{align}
where $r_h$ is the event horizon.

We are only interesting in the effect of MK parameter $\gamma$ on the chaotic dynamics. So we let $k=0$ through this paper.

\section{ Ring string around the conformal black hole}\label{sec-string}

Now, we consider the motion of a  ring string around the conformal black hole.
The  ring string can be depicted by the Polyakov action,
\begin{align}
\mathcal{L}=-\frac{1}{2\pi\alpha}\sqrt{-g}g^{\mu\nu}G_{ab}\partial_{\mu}X^{a}\partial_{\nu}X^{b}\,.
\label{L}
\end{align}
The Polyakov action is on-shell equivalent to the Nambu-Goto action.
$\alpha$ is the coupling constant relating the string length $l_s$ by $l_s^2=\alpha$.
$X^{a}$ is the coordinates of the target space and $G_{ab}$ the corresponding metric.
The world sheet of the string is described by the coordinates $\sigma^{\mu}=(\tau,\sigma)$ with the induced metric $g_{\mu\nu}$ on the world sheet. It is convenient to work in the conformal gauge $g_{\mu\nu}=\eta_{\mu\nu}$. And then, we take the following ansatz
\begin{eqnarray}
t=t(\tau),\ \ \  r=r(\tau),\ \ \ \theta=\theta(\tau),\ \ \ \phi=\eta\sigma\,.
\label{ansatz}
\end{eqnarray}
The winding number $\eta$ depicts the differences between strings and particles.
Under the above ansatz, the Polyakov Lagrangian can be explicitly worked out as
\begin{align}
\mathcal{L}=\frac{\dot{r}(\tau)^{2}}{2\pi\alpha B(r)}-\frac{B(r)\dot{t}(\tau)^{2}}{2\pi\alpha}
+\frac{r^{2}\dot{\theta}(\tau)^{2}}{2\pi\alpha}-\frac{r^{2}\eta^{2}\sin^{2}(\theta)}{2\pi\alpha}\,,
\label{L1}
\end{align}
where the dot denotes the derivative with respect to $\tau$.
The corresponding Hamiltonian is
\begin{align}
H=-\frac{P_{t}^{2}\pi\alpha}{2 B(r)}+\frac{1}{2}\pi\alpha B(r)P_{r}^{2}+\frac{\pi\alpha P_{\theta}^{2}}{2r^{2}}+\frac{r^{2}\eta^{2}\sin^{2}(\theta)}{2\pi\alpha}\,,
\label{H}
\end{align}
which satisfies the constraint $H=0$. $\{t,P_t\}$, $\{r,P_r\}$, $\{\theta,P_{\theta}\}$ are the canonical phase space variables with
\begin{eqnarray}
P_{t}=-\frac{B(r)\dot{t}(\tau)}{\pi\alpha},\ \ \  P_{r}=\frac{\dot{r}(\tau)}{\pi\alpha B(r)},\ \ \ P_{\theta}=\frac{r^{2}\dot{\theta}(\tau)}{\pi\alpha}\,.
\end{eqnarray}
Then the canonical equations of motion can be derived by the Poisson bracket,
\begin{eqnarray}
&&
\dot{t}=-\frac{\pi \alpha }{B(r)}P_t\,,
\label{eom-dott}
\
\\
&&\dot{r}=\pi \alpha B(r) P_r\,,
\label{eom-dotr}
\\
&&\dot{\theta}=\frac{\pi \alpha }{r^2}P_{\theta}\,,
\label{eom-dottheta}\\
&&\dot{P_t}=0\,,\label{eom-dotpt} \\
&&\dot{P_r}=\frac{\pi \alpha }{r^3}P_{\theta}^2 - \frac{\pi \alpha B'(r)}{2B(r)^2}P_t^2-\frac{1}{2}\pi \alpha B'(r) P_r^2-\frac{r \eta^2 \sin^{2}(\theta)}{\pi \alpha}\,,
\label{eom-dotpr}
\
\\
&&
\dot{P_\theta}=-\frac{r^{2}\eta^{2}\sin(\theta)\cos(\theta)}{\pi\alpha}\,,
\label{eom-dotptheta}
\end{eqnarray}
where the prime represents the derivative with respective to $r$.
Eq.\eqref{eom-dotpt} gives a constant of motion $P_{t}=E$, which relates to the energy.

\section{Chaotic dynamics of string around the conformal black hole}\label{sec-chaodyn}
\subsection{Numerical method}

In this section, we shall numerically solve the dynamical system of  ring string around the conformal black hole described above.
Most nonlinear systems are not integrable, in order to get a better approximate solution, numerical methods are needed.
High precision numerical solution is crucial to the chaotic system,
the reason for this is that the low precision numerical solutions will produce pseudo chaos.
It is generally known that forth-order Runge-Kutta algorithm(RK4) is very effective to deal with the ordinary differential equations,
RK4 has the advantages of symmetrical structure, low calculation amount and convenient to use.
But RK4 is not suitable for long-term integration because of the accumulation of truncation error.
As was stated in \cite{Ma:2019ewq}, the original hyper-surface will be deviated by RK4.
Fortunately, it has been reported that the velocity scaling method \cite{NewAMa} is very useful to treat such problems.
Thanks to its strict restraint mechanism, it provides great control upon the output accuracy.
The accuracy can be guaranteed at any time by evaluating constraint ($H=0$).
The constraint represents the energy conservation condition for the
motion of circular ring in the charged black hole background.
The effectiveness of the velocity scaling method has been verified by large numbers of numerical experiments.
For more details, please see the report in \cite{NewAMa}.

Although the time evolution can provide a visual picture of dynamics of a ring string, it is not worth promoting.
The reason for this is that the result is not very effective in the long-term integration.
Therefore, in order to get a better perspective of the ring string dynamics, we need other ways, such as chaos indicators.

Effective chaos indicator is very important to discuss the evolution of the chaotic system.
As the chaotic system is highly sensitive to their initial conditions,
it generates a large number of chaos indicators.
Such as spectrum analysis, bifurcations, fractal theory, Poincare sections, Lyapunov exponent,
fast Lyapunov indicator, relative finite-time Lyapunov indicator, smaller alignment index, and generalized alignment index, etc.
There is a well-documented discussion of the characteristics of these chaos indicators in the reference \cite{MaIJBC},
we won't explore it in this paper.
It has been proved that Lyapunov exponent is not only useful in conservative systems, but also is effective in dissipative systems \cite{Ma:2019ewq,MaIJBC}. According to this,
Lyapunov exponent will be used here to help us explore the dynamics of string motion with Weyl conformal gravity.

As a popular and powerful chaos indicator,
Lyapunov exponent determining whether an orbit is chaotic or not by means of
measure the average rate of divergence of two nearby trajectories in phase space.
There are two ways to calculate the Lyapunov exponent, the one is the variational method, the other is the two particle method.
The difference between the two methods is that the former integrates both the motion and variational equations at the same time.
but the later integrates the motion equation twice with two nearby orbit initial values.
For an n-dimensional system, it has n Lyapunov exponents.
As long as one of them is greater than zero, the system is chaotic.
However, the Lyapunov exponent is always replaced by the maximum Lyapunov exponent.
That is because all Lyapunov directions will converge to one tangent vector if the Gram-Schmidt orthogonalization \cite{Benettin1976}
has not been considered at every step.
The form of the maximum Lyapunov exponent is
\begin{eqnarray}
\label{constant}
\lambda = \lim_{\tau\rightarrow \infty}\frac{1}{\tau}\ln\frac{\|\bm{\xi}(\tau)\|}{\|\bm{\xi}(0)\|}.
\end{eqnarray}
$\bm{\xi}(0)$ and $\bm{\xi}(\tau)$ denote the distances at the starting point and time $\tau$.
If $\lambda > 0$, that is the bounded orbit is chaotic.
But for $\lambda = 0$, the orbit is order. It is more effective to use $log-log$ plot to describe the dynamics. In this convention, the motion is ordered if $log_{10}|\lambda|$ decreases linearly with $log_{10}(\tau)$ increasing, while the motion is chaotic if $log_{10}|\lambda|$ exponentially changes with $log_{10}(\tau)$. In this paper, we shall use this formula to depict the dynamics of the string.

\begin{figure}
	\includegraphics[width=0.45\textwidth]{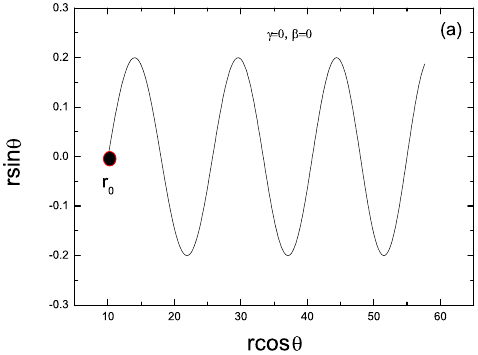}
    \includegraphics[width=0.45\textwidth]{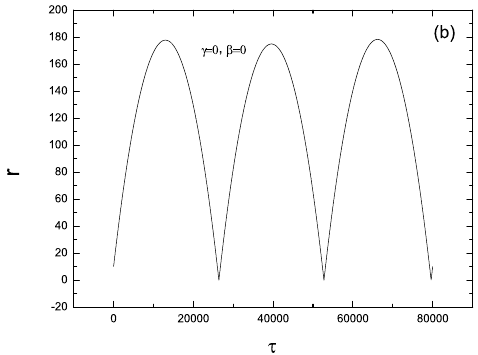}
    \caption{The string trajectory around the massive body with $\gamma=0$ and $\beta=0$, for which $B(r)=1$ is independent of $r$ and the horizon is absent.
    Here, we set $E=12$, $\alpha=1/\pi$ and the initial conditions as $r_{0}= 10$, $\theta_{0}=0$, $p_{\theta_{0}}=2.5679$.}
    \label{fig-evo-v1}
\end{figure}
\begin{figure}
	\includegraphics[width=0.45\textwidth]{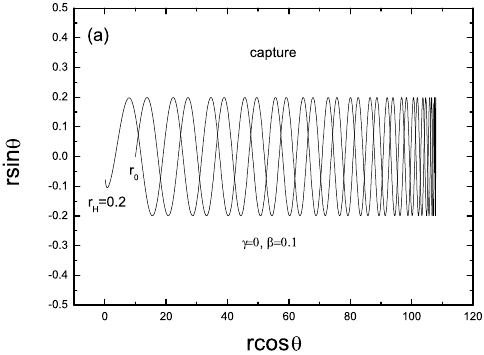}
    \includegraphics[width=0.45\textwidth]{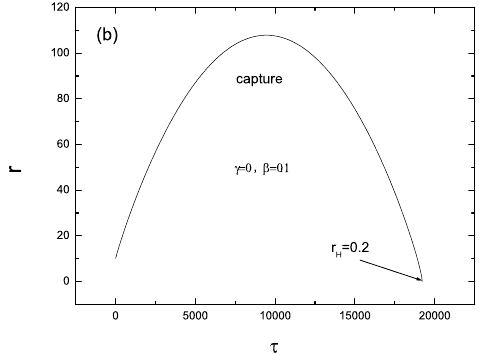}
    \caption{The string trajectory around the Schwarzschild black hole $\beta=0.1$. Here, we set $E=12$, $\alpha=1/\pi$ and the initial conditions as $r_{0}= 10$, $\theta_{0}=0$, $p_{\theta_{0}}=2.5679$.
    }
    \label{fig-evo-v2}
\end{figure}

Before proceeding to explore the chaotic dynamics of the string in detail, we want to gain a visual picture of dynamics of the string by studying the time evolution of $R(\tau)$.
To this end, we can directly solve the canonical equations of motion (Eq. \eqref{eom-dott} to Eq.\eqref{eom-dotptheta}) by the velocity scaling method.

Fig.\ref{fig-evo-v1} and Fig.\ref{fig-evo-v2} exhibit the string trajectory around the black hole with two simple examples. One is $B(r)=1$ and another is the Schwarzschild black hole with $\beta=0.1$. Here, we set $E=12$, $\alpha=1/\pi$ and the initial conditions as $r_{0}= 10$, $\theta_{0}=0$, $p_{\theta_{0}}=2.5679$ without loss of generality.

From Fig.\ref{fig-evo-v1}, we see that the string always oscillates back and forth around the black hole. The string is in a simple harmonic vibration and the motion is ordered. For the case of the Schwarzschild black hole with $\beta=0.1$, we find that after a finite number of oscillations, the string is captured by the black hole (Fig.\ref{fig-evo-v2}).

\begin{figure}
	\includegraphics[width=0.5\textwidth]{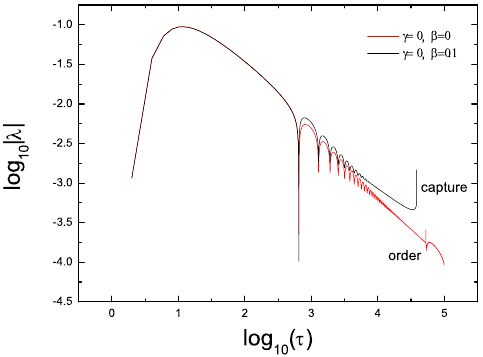}
    \caption{The corresponding maximum Lyapunov exponents of the string trajectory shown in Fig.\ref{fig-evo-v1} and Fig.\ref{fig-evo-v2}.}
    \label{fig-evo-v4}
\end{figure}

Just as pointed out above, although the string trajectory can provide a visual picture, the calculation is time-consuming and inefficient.
It is more efficient to study the chaotic dynamics of the string by the Lyapunov exponent.
Here, we use the two particle method, so that there is no need to compute the variational equation. The initial distance between the two nearby orbit is $10^{-8}$.
For comparison, we calculate the corresponding maximum Lyapunov exponents for Fig.\ref{fig-evo-v1} and Fig.\ref{fig-evo-v2}, which are shown in Fig.\ref{fig-evo-v4}.
We find that both curves are almost the same at the initial phase of the evolution.
It means that there is a finite number of oscillations around the black hole.
As time goes, the string exhibits different dynamical behaviors for different parameters.
When $\gamma=0$ and $\beta=0$, which is just flat space, we see that $log_{10}|\lambda|$ decreases linearly with $log_{10}(\tau)$ increasing (red curve in Fig.\ref{fig-evo-v4}). It indicates that the orbit is ordered.
But for $\gamma=0$ and $\beta=0.1$, which is the case of Schwarzschild black hole, the curve exhibits different features (black curve in Fig.\ref{fig-evo-v4}).
After undergoing long period of oscillation,
the system automatically terminates the calculation in the later stage.
The accident is due to the distance between the two adjacent orbits becomes too short with time.
That is, the orbit motion is collapsed, and the string is captured by the black hole.
That is to say, $log_{10}|\lambda|$ changes non-linearly with $log_{10}(\tau)$. Therefore, the orbit motion is chaotic.

As stated in \cite{Ma:2019ewq}, in order to intuitively observe the difference between different cases, the chaos indicators are strongly recommended, not the trajectory pictures.
The principal reason is that it is hard to tell the difference between the order orbit and the chaotic orbit in the long term integration.
However, the chaos indicators are the powerful ways to deal with these problems.
As shown above, the method of the maximum Lyapunove exponent exhibits the efficient power to detect the chaotic effects. So we shall mainly use the maximum Lyapunove exponent to study the chaotic behavior in this paper.

\subsection{Chaotic dynamics over Schwarzschild black hole}

\begin{figure}
	\includegraphics[width=0.45\textwidth]{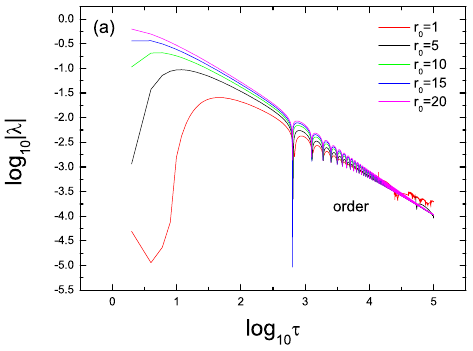}\hspace{1cm}
	\includegraphics[width=0.45\textwidth]{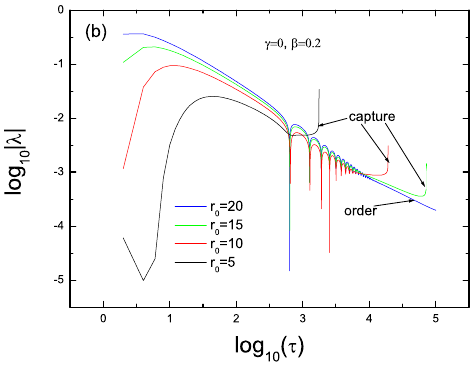}
    \caption{Left plot: The maximum Lyapunov exponents of the string motion with different $r_0$ in flat space ($\gamma=0$ and $\beta=0$). Right plot: The maximum Lyapunov exponents of the string motion with different $r_0$ over the asymptotically flat Schwarzschild black hole ($\gamma=0$ and $\beta=0.2$).}
    \label{fig-evo-v6}
\end{figure}
\begin{widetext}
	\begin{table}[ht]
		\begin{center}
			\begin{tabular}{|c|c|c|c|c|c|c|c|c|c|c|c|c|c|c|c|c|c|c|c|c|c|c|c|}
				\hline
				~$r_0$~&~$1$&~$2$~&~$3$~&~$4$&~$5$~&~$6$&~$7$~&~$8$~&~$9$&~$10$~&~$11$&~$12$~&~$13$~&~$14$&~$15$~&~$16$&~$17$~&~$18$~&~$19$&~$20$
				\\
				\hline
				~$\beta=0$~ & ~$O$~& ~$O$~&~$O$~&~$O$&~$O$~ & ~$O$~& ~$O$~&~$O$~&~$O$&~$O$~ & ~$O$~& ~$O$~&~$O$~&~$O$&~$O$~ & ~$O$~& ~$O$~&~$O$~&~$O$&~$O$
				\\
				\hline
				~$\beta=0.2$~ &~$C$~& ~$C$~&~$C$~&~$C$&~$C$~ &~$C$~& ~$C$~&~$C$~&~$C$&~$C$~ &~$C$~& ~$C$~&~$C$~&~$C$&~$C$~ &~$O$~& ~$O$~&~$O$~&~$O$&~$O$
				\\
				\hline
			\end{tabular}
			\caption{\label{table-v1} The dynamics behaviors of the string with $\beta=0$ and $\beta=0.2$ for sample $r_0$ ($\gamma=0$). ``O" and ``C" denote ``Ordered" and ``Chaotic and being captured", respectively. These results are given by observing the behaviors of the maximum Lyapunov exponent $log_{10}|\lambda|$ as the function of $log_{10}(\tau)$.}
		\end{center}
	\end{table}
\end{widetext}

When $\gamma=0$, the conformal black hole reduces to the Schwarzschild black hole. There is a special case of $\beta=0$, for which the Schwarzschild solution is flat. When $\beta\neq 0$, the Schwarzschild black hole has an event horizon. We will further make an exploration on the dynamics of motion over Schwarzschild black hole.

In \cite{Frolov:1999pj}, the authors explore the different types of trajectories and the fractal dimension as a function of energy, by which they discuss transition from order to chaos.
For completeness and making comparisons to the chaotic dynamics over conformal black hole, here we reexamine the chaotic dynamics of the string over Schwarzschild black hole by the maximum Lyapunov exponent described in the above subsection.

When $\beta=0$, the space is just flat. We show the maximum Lyapunov exponent $log_{10}|\lambda|$ as the function of $log_{10}(\tau)$ for different initial positions $r_0$ in left plot of Fig.\ref{fig-evo-v6}. It is obvious that the behaviors are ordered, i.e., integrable. Further, we summary the behaviors of the string with $\beta=0$ for more $r_0$ in Table \ref{table-v1}. We find that indeed in the flat space, the motion of the string is ordered, which is independent of the initial position of the string.

Then, we turn to study the case of $\beta=0.2$, which is a Schwarzschild black hole with the horizon locating at $r_h=0.4$. We show the maximum Lyapunov exponent $log_{10}|\lambda|$ as the function of $log_{10}(\tau)$ for different initial positions $r_0$ in the right plot of Fig.\ref{fig-evo-v6} and also summary the behaviors of the string with $\beta=0.2$ for more $r_0$ in Table \ref{table-v1}.
We find that when the initial position of the string is far away from the black hole ($r_{0}\geq 16$), the string oscillates around the black hole and its motion is ordered.
As the initial position of the string approaches the black hole, the motion of the string is chaotic and finally the string is captured by the black hole.

All the observations here by the maximum Lyapunov exponents are in agreement with that in \cite{Frolov:1999pj}.

\subsection{Chaotic dynamics around conformal black hole}

Now, we turn to study the chaotic dynamics around the conformal black hole with general MK parameter $\gamma$.
It is convenient to work with dimensionless parameters. So we make the following rescaling
\fa
\label{rescaling}
ds\rightarrow \beta ds\,,~~~~~~t\rightarrow\beta t\,,~~~~~~r\rightarrow\beta r\,,~~~~~~
\gamma\rightarrow\frac{\gamma}{\beta}\,.
\ffa
Under this rescaling, we can set $\beta=1$ in what follows.
Depending on the parameters, there are rich horizon structures (see for example Refs.\cite{Abbas:2020pzc,Villanueva:2013gga,Turner:2020gxo} for detailed discussions):
\begin{itemize}
  \item \textbf{Case I}: No horizon for $\gamma<-1/3$ and $\gamma>1$. But when $\gamma<-1/3$, $B(r)<0$, for which we don't consider here.
  \item \textbf{Case II}: Only the black hole event horizon for $0<\gamma<2/3$.
  \item \textbf{Case III}: Two horizons including black hole event horizon and Cauchy horizon for $2/3<\gamma<1$.
  \item \textbf{Case IV}: Two horizons including black hole event horizon and cosmological horizon for $-1/3<\gamma<0$.
\end{itemize}

\begin{figure}
	\includegraphics[width=0.48\textwidth]{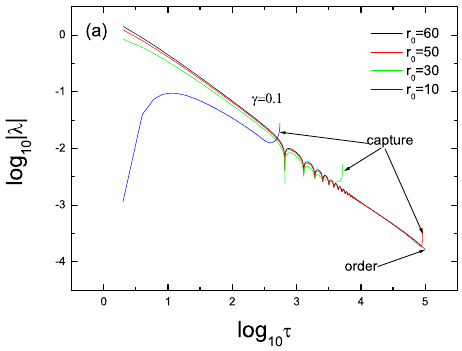}\hspace{0.5cm}
	\includegraphics[width=0.48\textwidth]{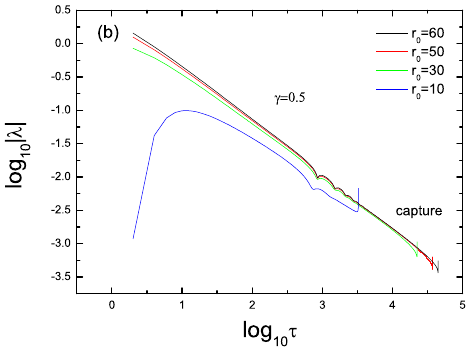}
	\includegraphics[width=0.48\textwidth]{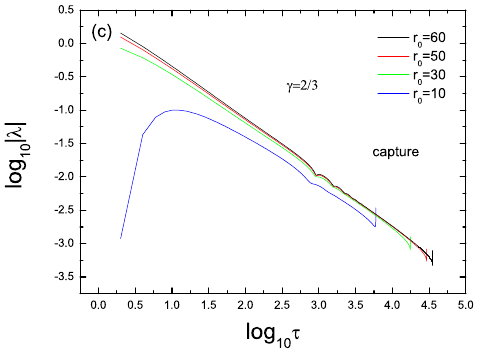}\hspace{0.5cm}
	\includegraphics[width=0.48\textwidth]{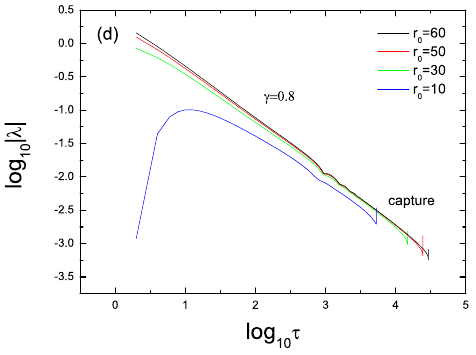}
	\includegraphics[width=0.48\textwidth]{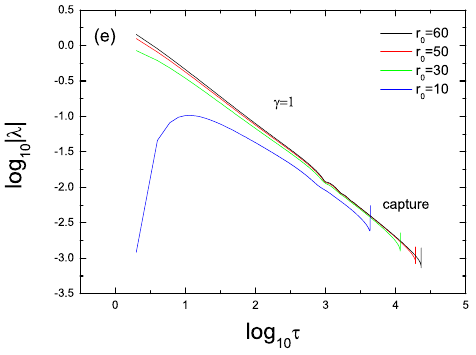}\hspace{0.5cm}
	\includegraphics[width=0.48\textwidth]{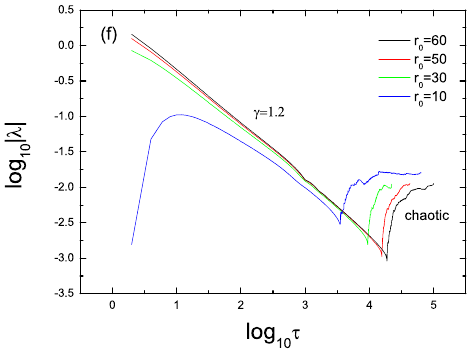}
	\caption{The maximum Lyapunov exponents of the string trajectory for fixed $\gamma$ with different initial position $r_{0}=10, 30, 50, 60$.}
	\label{fig6}
\end{figure}

\begin{figure}
	\includegraphics[width=0.48\textwidth]{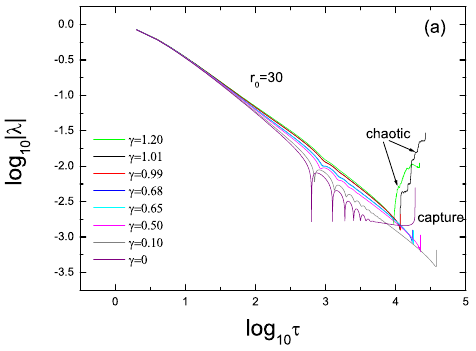}\hspace{0.5cm}
	\includegraphics[width=0.48\textwidth]{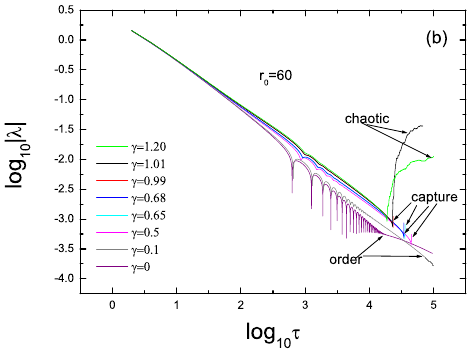}
	\caption{The maximum Lyapunov exponents of the string trajectory for different $\gamma$ with fixed $r_0$ (left pannel is for $r_0$=30 and right pannel for $r_0=60$)}
	\label{fig7}
\end{figure}

The initial position $r_0$ and the MK parameter $\gamma$ are the key ingredients affecting the chaotic dynamics. 
We show the maximum Lyapunov exponents of the string trajectory for fixed $\gamma$ with different initial position of the string in 
Fig.\ref{fig6}. To more clearly see the effect of the parameter $\gamma$, we also show the maximum Lyapunov exponents for chosen initial position with different $\gamma$ in Fig.\ref{fig7} and summary the dynamics behaviors of the string in Table \ref{table-v2}.
In what follows, we shall discuss the main dynamics characteristics of the string.

For $\gamma$ in the region of $0\leq\gamma\leq 0.1$, we find that when the string is placed close to the black hole at the beginning, the motion of the string is chaotic and finially the string is captured by the black hole. As the initial position $r_0$ increases, the captured time also increases. Further increasing $r_0$, we observe that the system becomes ordered (see pannel (a) in Fig.\ref{fig6}, right pannel in Fig.\ref{fig7} and Table \ref{table-v2}).

Then, by lots of numerical simulations, we find that when $\gamma$ in the region of $0.1<\gamma\leq 1$, the motion of the string is chaotic and finally the string is captured by the black hole even for the initial position being far away from the horizon of the black hole. We present sample $\gamma$ in the pannels (b), (c), (d), (e) in Fig.\ref{fig6} (also see Fig.\ref{fig7} and Table \ref{table-v2}).

Notice that the captured time doesn't linearly change with the MK parameter $\gamma$. When $\gamma\leq 0.1$, it is easier for the string to be captured by the black hole with the increase of $\gamma$. But when $\gamma=0.5$, the captured time almost reaches the maximum value. And then, further increasing $\gamma$, the captured time decreases, which means that it is easier for the string to be captured by the black hole. But on the whole, it is easier for the string to be captured by the black hole for the $\gamma$ being in the region of $\gamma\leq 0.1$ than that in the region of $\gamma\geq0.5$.

We are particularly interested in if any sharp transition in chaotic dynamics of the string happens when geometry changes qualitatively.
To this end, we explore the case for $\gamma=2/3$, which the critical value dividing the black hole with one horizon (Case II) between the black hole with two horizons, and we cannot find any sharp transition in chaotic dynamics happening. However, once the system develops into a massive body without horizon (Case I for $\gamma>1$), the motion of the stirng is chaotic but is not captured by the massive body (see pannel (f) in Fig.\ref{fig6}, Fig.\ref{fig7} and Table \ref{table-v2}). Therefore, we conclude that there a sharp transition in chaotic dynamics when the horizon disappears.

\begin{widetext}
	\begin{table}[ht]
		\begin{center}
			\begin{tabular}{|c|c|c|c|c|c|c|c|c|c|c|c|c|}
				\hline
				~$\gamma$~&~$0.1$&~$0.2$~&~$0.3$~&~$0.4$&~$0.5$~&~$0.6$&~$0.7$~&~$0.8$~&~$0.9$&~$1.0$~&~$1.1$&~$1.2$
				\\
				\hline
				~$r_{0}=10$~ & ~$C$~& ~$C$~&~$C$~&~$C$&~$C$~ & ~$C$~& ~$C$~&~$C$~&~$C$&~$C$~ & ~$\slashed{C}$~& ~$\slashed{C}$
				\\
				\hline
				~$r_{0}=30$~ & ~$C$~& ~$C$~&~$C$~&~$C$&~$C$~ & ~$C$~& ~$C$~&~$C$~&~$C$&~$C$~ & ~$\slashed{C}$~& ~$\slashed{C}$
				\\
				\hline
				~$r_{0}=50$~ & ~$C$~& ~$C$~&~$C$~&~$C$&~$C$~ & ~$C$~& ~$C$~&~$C$~&~$C$&~$C$~ & ~$\slashed{C}$~& ~$\slashed{C}$
				\\
				\hline
				~$r_{0}=60$~ & ~$O$~& ~$C$~&~$C$~&~$C$&~$C$~ & ~$C$~& ~$C$~&~$C$~&~$C$&~$C$~ & ~$\slashed{C}$~& ~$\slashed{C}$
				\\
				\hline
			\end{tabular}
			\caption{\label{table-v2} The dynamics behaviors of the string for different $\gamma$ and $r_0$. ``O", ``C" and ``$\slashed{C}$" denote ``Ordered", ``Chaotic and being captured" and ``Chaotic but not being captured", respectively. These results are given by observing the behaviors of the maximum Lyapunov exponent $log_{10}|\lambda|$ as the function of $log_{10}(\tau)$.}
		\end{center}
	\end{table}
\end{widetext}

Finally, we present a simple discussion on the case IV ($-1/3<\gamma<0$), for which the black hole has two horizons including black hole event horizon and cosmological horizon. For this case, since we have to consider the boundary condition at the cosmological horizon if the string touches it, the calculation becomes subtle. Fortunately, by studying the maximum Lyapunov exponents of the ring string trajectory with different $\gamma$, we find that as long as the initial position of the string doesn't approach the cosmological horizon, the strings are all captured by the black hole such that they doesn't touch the cosmological horizon. Indeed, if we place the sting approaching the cosmological horizon at the beginning, it is possible for the string to touch the cosmological horizon and then the calculation becomes subtle. For this case, we leave for future study.

\subsection{Chaos bound}

Recently a great advance on chaotic dynamics is that there is a universal upper bound of Lyapunov exponent, which is also dubbed as MSS bound, in quantum field theories \cite{Maldacena:2015waa}. In quantum field theory, we can calculate the out-of-time-correlation function (OTOC) to extract the Lyapunov exponent, which is expected as \cite{Kitaev:2014}
\fa
\lambda\leq 2\pi T\,,
\label{lambdabound}
\ffa
where $T$ is the temperature of the system. Here, we also call the above inequality as the MSS inequality. The bound is saturated by the hologaphic dual field theory. It is expected because the black hole is the faster scrambler \cite{Sekino:2008he}. An particular interesting development is that the Sachdev-Ye-Kitaev (SYK) model \cite{Sachdev:1992,Kitaev:2015} also saturates the MSS bound, which bridges physics of black hole and condensed matter theory (see for example \cite{Sachdev:2015efa,Stanford:2015owe,Fu:2016yrv,Berenstein:2015yxu}).

On the other hand, the authors in \cite{Hashimoto:2016dfz} study the chaotic dynamics of paticle over an AdS black hole and they find that the Lyapunov exponent of particle motion is also subject to the inequality \eqref{lambdabound}.
Further, \v{C}ubrovi\'c studies the chaotic dynanics of closed strings in AdS black hole background and they find the following generalized MSS inequality \cite{Cubrovic:2019qee}
\fa
\label{boundstring}
\lambda\leq 2\pi T \eta\,,
\ffa
holds. Recalling that $\eta$ is the winding number of the string in Eq.\eqref{ansatz}. When the particle or string moves near the black hole horizon, the Lyapunov exponent $\lambda$ closely approaches the upper bound in Eq.\eqref{lambdabound} or Eq.\eqref{boundstring} \cite{Hashimoto:2016dfz,Cubrovic:2019qee}.

\begin{figure}
	\includegraphics[width=0.7\textwidth]{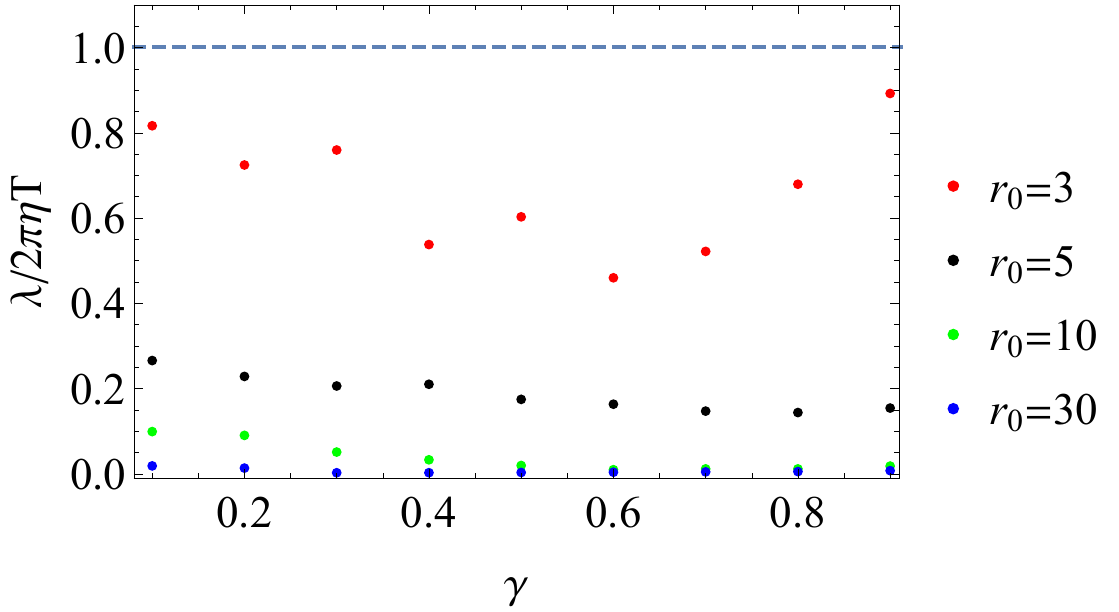}
	\caption{The maximum Lyapunov exponent $\lambda/2\pi\eta T$ for different $\gamma$ with various $r_0$. The dotted line is the MSS bound. 
}
	\label{figmssbound}
\end{figure}

Here we would like to examine if the above generalized MSS inequality \eqref{boundstring} still holds in the asymptotically flat black hole background. To this end, we show the maximum Lyapunov exponents $\lambda/2\pi\eta T$ for different $\gamma$ with various $r_0$ in Fig.\ref{figmssbound}. It is obvious that the generalized MSS inequality holds. Especially, we observe that as the initial position of the string approaches the black hole horizon, the Lyapunov exponent $\lambda$ also approaches the upper bound in Eq.\eqref{boundstring}. This observation is similar to that in AdS black hole background studied in \cite{Cubrovic:2019qee}. Therefore, we infer that the inequalities \eqref{lambdabound} and \eqref{boundstring} hold not only in AdS black hole background but also in asymptotically flat balck hole background. In future, we shall further examine and prove this observation by numerical and analytical methods.

\section{Conclusions and discussions}\label{sec-con}

In this paper, by calculating the maximum Lyapunov exponent, we explore the chaotic dynamics of the string around the conformal black hole.
We mainly study the effect of the characteristic parameter $\gamma$ of the conformal black hole on the chaotic behaviors. We summary the main properties of the chaotic behaviors as what follows.
\begin{itemize}
	\item When $\gamma$ is in the region of $0\leq\gamma\leq 0.1$, the chaotic behavior heavily depends on the initial position $r_0$ of the string. When the initial position of string approaches to the black hole at the beginning, the motion of the string is chaotic and finially the string is captured by the black hole. Notice that the capture time also increases with $r_0$ increasing. As $r_0$ further increases, the system becomes ordered.
	\item When $\gamma$ in the region of $0.1<\gamma\leq 1$, the motion of the string is chaotic and finally the string is captured by the black hole even for the initial position being far away from the horizon of the black hole.
	\item There is a sharp transition in chaotic dynamics when the horizon disappears. To be more specific, the motion of the string is chaotic and finially the string is captured by the black hole \footnote{In the region of $0\leq\gamma\leq 0.1$, the motion of the string can be ordered if the string is placed far away from the black hole.}. However, once the system develops into a massive body without horion, the motion of the string is chaotic but the string is not captured.
	\item The generalized MSS inequality holds even in an asymptotically flat black hole background. Especially, as the initial position of the string approaches the black hole horizon, the Lyapunov exponent also approaches the upper bound of the generalized MSS inequality.
	\item When the cosmological horizon is included, the calculation becomes subtle. But for the case of $-1/3<\gamma<0$ studied here, we find that as long as the initial position of the string doesn't approach the cosmological horizon, the strings are all captured by the black hole such that they doesn't touch the cosmological horizon.
\end{itemize}

In this paper, we let $k=0$ and only focus on the effect of the MK parameter $\gamma$. In future, we shall study the joint effect of $\gamma$ and $k$. For $k>0$, the black hole shall include the cosmological horizon, we need develop a new method to calculate the maximum Lyapunove exponent to study the chaotic behavior. Another interesting case is for $k<0$, which corresponds an asymptotically AdS spacetime. It is interesting to study the dual interpretation of the ring string around the conformal black hole.

\begin{acknowledgments}

This work is supported by the Natural Science
Foundation of China under Grant Nos. 12073008, 11703005, 11775036, 12147209.
Da-Zhu Ma is supported by the Young Top-notch Talent Cultivation Program of Hubei Province.
Guoyang Fu is supported by the Postgraduate Research \& Practice Innovation Program of Jiangsu Province (KYCX20\_2973).
Jian-Pin Wu is also supported by Top Talent Support Program from Yangzhou University.

\end{acknowledgments}



\begin{thebibliography}{99}



\bibitem{Carter:1968}
B. Carter, ``Global structure of the Kerr family of gravitational fields,'' Phys. Rev. \textbf{174} (5): 1559 C1571 (1968).

\bibitem{Dettmann:1994dj}
  C.~P.~Dettmann, N.~E.~Frankel and N.~J.~Cornish,
  ``Fractal basins and chaotic trajectories in multi - black hole space-times,''
  Phys.\ Rev.\ D {\bf 50}, R618 (1994)
  [gr-qc/9402027].

\bibitem{Hanan:2006uf}
  W.~Hanan and E.~Radu,
  ``Chaotic motion in multi-black hole spacetimes and holographic screens,''
  Mod.\ Phys.\ Lett.\ A {\bf 22}, 399 (2007)
  [gr-qc/0610119].

\bibitem{Karas:1992}
V. Karas and D. Vokrouhlicky, ``Chaotic motion of test particles in the Ernst space-time,''
Gen. Relativ. Gravit. \textbf{24} (1992) 729.

\bibitem{Bombelli:1992}
L. Bombelli and E. Calzetta, ``Chaos around a black hole,''
Class. Quant. Grav. \textbf{9} (1992) 2573.

\bibitem{Aguirregabiria:1996vq}
  J.~M.~Aguirregabiria,
  ``Chaotic scattering around black holes,''
  Phys.\ Lett.\ A {\bf 224}, 234 (1997)
  [gr-qc/9604032].

\bibitem{Sota:1995ms}
  Y.~Sota, S.~Suzuki and K.~i.~Maeda,
  ``Chaos in static axisymmetric space-times. 1: Vacuum case,''
  Class.\ Quant.\ Grav.\  {\bf 13}, 1241 (1996)
  [gr-qc/9505036].

\bibitem{Chen:2016tmr}
  S.~Chen, M.~Wang and J.~Jing,
  ``Chaotic motion of particles in the accelerating and rotating black holes spacetime,''
  JHEP {\bf 1609}, 082 (2016)
  [arXiv:1604.02785 [gr-qc]].

\bibitem{Varvoglis:1992}
H. Varvoglis and D. Papadopoulos, ``Chaotic interaction of charged particles with a gravitational wave,''
Astron. Astrophys. \textbf{261} (1992) 664.


\bibitem{Frolov:1999pj}
  A.~V.~Frolov and A.~L.~Larsen,
  ``Chaotic scattering and capture of strings by black hole,''
  Class.\ Quant.\ Grav.\  {\bf 16}, 3717 (1999)
  [gr-qc/9908039].

\bibitem{Zayas:2010fs}
  L.~A.~Pando Zayas and C.~A.~Terrero-Escalante,
  ``Chaos in the Gauge / Gravity Correspondence,''
  JHEP {\bf 1009}, 094 (2010)
  [arXiv:1007.0277 [hep-th]].


\bibitem{Ma:2014aha}
  D.~Z.~Ma, J.~P.~Wu and J.~Zhang,
  ``Chaos from the ring string in a Gauss-Bonnet black hole in AdS5 space,''
  Phys.\ Rev.\ D {\bf 89}, no. 8, 086011 (2014)
  [arXiv:1405.3563 [hep-th]].

\bibitem{Bai:2014wpa}
  X.~Bai, B.~H.~Lee, T.~Moon and J.~Chen,
  ``Chaos in Lifshitz Spacetimes,''
  J.\ Korean Phys.\ Soc.\  {\bf 68}, no. 5, 639 (2016)
  [arXiv:1406.5816 [hep-th]].

\bibitem{Basu:2016zkr}
  P.~Basu, P.~Chaturvedi and P.~Samantray,
  ``Chaotic dynamics of strings in charged black hole backgrounds,''
  Phys.\ Rev.\ D {\bf 95}, no. 6, 066014 (2017)
  [arXiv:1607.04466 [hep-th]].

\bibitem{Ishii:2016rlk}
  T.~Ishii, K.~Murata and K.~Yoshida,
  ``Fate of chaotic strings in a confining geometry,''
  Phys.\ Rev.\ D {\bf 95}, no. 6, 066019 (2017)
  [arXiv:1610.05833 [hep-th]].

\bibitem{Cubrovic:2019qee}
M.~\v{C}ubrovi\'c,
``The bound on chaos for closed strings in Anti-de Sitter black hole backgrounds,''
JHEP \textbf{12}, 150 (2019)
[arXiv:1904.06295 [hep-th]].

\bibitem{Ma:2019ewq}
D.~Z.~Ma, D.~Zhang, G.~Fu and J.~P.~Wu,
``Chaotic dynamics of string around charged black brane with hyperscaling violation,''
JHEP \textbf{01} (2020), 103
[arXiv:1911.09913 [hep-th]].

\bibitem{Vilenkin:1994}
A.~Vilenkin and E. P. S. Shellard, ``Cosmic Strings and other Topological Defects,'' 1994,
Cambridge University Press, Cambridge.


\bibitem{Maldacena:1997re}
  J.~M.~Maldacena,
  ``The Large N limit of superconformal field theories and supergravity,''
  Int.\ J.\ Theor.\ Phys.\  {\bf 38}, 1113 (1999)
  [Adv.\ Theor.\ Math.\ Phys.\  {\bf 2}, 231 (1998)].
  [hep-th/9711200].

\bibitem{Gubser:1998bc}
  S.~S.~Gubser, I.~R.~Klebanov and A.~M.~Polyakov,
  ``Gauge theory correlators from noncritical string theory,''
  Phys.\ Lett.\ B {\bf 428}, 105 (1998).
  [hep-th/9802109].

\bibitem{Witten:1998qj}
  E.~Witten,
  ``Anti-de Sitter space and holography,''
  Adv.\ Theor.\ Math.\ Phys.\  {\bf 2}, 253 (1998).
  [hep-th/9802150].

\bibitem{Aharony:1999ti}
  O.~Aharony, S.~S.~Gubser, J.~M.~Maldacena, H.~Ooguri and Y.~Oz,
  ``Large N field theories, string theory and gravity,''
  Phys.\ Rept.\  {\bf 323}, 183 (2000).
  [hep-th/9905111].

\bibitem{Weyl:1918pdp}
H.~Weyl,
``Reine Infinitesimalgeometrie,''
Math. Z. \textbf{2}, no.3-4, 384-411 (1918).

\bibitem{MFathi:2019jgd}
M.~Fathi, M.~Olivares and J.~R.~Villanueva,
``Classical tests on a charged Weyl black hole: bending of light, Shapiro delay and Sagnac effect,''
Eur. Phys. J. C \textbf{80}, no.1, 51 (2020)
[arXiv:1910.12811 [gr-qc]].

\bibitem{Mannheim:1988dj}
P.~D.~Mannheim and D.~Kazanas,
``Exact Vacuum Solution to Conformal Weyl Gravity and Galactic Rotation Curves,''
Astrophys. J. \textbf{342}, 635-638 (1989).

\bibitem{Mannheim:2006}
 P.D. Mannheim, ``Alternatives to dark matter and dark energy,''
 Prog. Part. Nucl. Phys. 56, 340 (2006).

\bibitem{Nesbet:2013}
 R.K. Nesbet, ``Conformal Gravity: Dark Matter and Dark Energy,''
 Entropy 15, 162 (2013).

\bibitem{Bambi:2016yne}
C.~Bambi, L.~Modesto, S.~Porey and L.~Rachwa\l{},
``Black hole evaporation in conformal gravity,''
JCAP \textbf{09}, 033 (2017)
[arXiv:1611.05582 [gr-qc]].

\bibitem{Kasikci:2018mtg}
O.~Ka\c{s}\i{}k\c{c}\i{} and C.~Deliduman,
``Gravitational Lensing in Weyl Gravity,''
Phys. Rev. D \textbf{100}, no.2, 024019 (2019)
[arXiv:1812.01076 [gr-qc]].

\bibitem{Fathi:2019jgd}
M.~Fathi, M.~Olivares and J.~R.~Villanueva,
``Classical tests on a charged Weyl black hole: bending of light, Shapiro delay and Sagnac effect,''
Eur. Phys. J. C \textbf{80}, no.1, 51 (2020)
[arXiv:1910.12811 [gr-qc]].

\bibitem{Takizawa:2020dja}
K.~Takizawa, T.~Ono and H.~Asada,
``Gravitational lens without asymptotic flatness: Its application to the Weyl gravity,''
Phys. Rev. D \textbf{102}, no.6, 064060 (2020)
[arXiv:2006.00682 [gr-qc]].

\bibitem{Li:2020wvn}
Z.~Li, G.~Zhang and A.~\"Ovg\"un,
``Circular Orbit of a Particle and Weak Gravitational Lensing,''
Phys. Rev. D \textbf{101}, no.12, 124058 (2020)
[arXiv:2006.13047 [gr-qc]].

\bibitem{Fathi:2020sey}
M.~Fathi, M.~Kariminezhad, M.~Olivares and J.~R.~Villanueva,
``Motion of massive particles around a charged Weyl black hole and the geodetic precession of orbiting gyroscopes,''
Eur. Phys. J. C \textbf{80}, no.5, 377 (2020)
[arXiv:2009.03399 [gr-qc]].

\bibitem{Fathi:2020sfw}
M.~Fathi, M.~Olivares and J.~R.~Villanueva,
``Gravitational Rutherford scattering of electrically charged particles from a charged Weyl black hole,''
Eur. Phys. J. Plus \textbf{136}, no.4, 420 (2021)
[arXiv:2009.03404 [gr-qc]].

\bibitem{Abbas:2020pzc}
G.~Abbas, M.~Azam and A.~Ditta,
``Accretion onto a born-Infeld black hole,''
Chin. J. Phys. \textbf{69}, 143-152 (2021)
[arXiv:2012.12035 [gr-qc]].

\bibitem{Konoplya:2020fwg}
R.~A.~Konoplya,
``Conformal Weyl gravity via two stages of quasinormal ringing and late-time behavior,''
Phys. Rev. D \textbf{103}, no.4, 044033 (2021)
[arXiv:2012.13020 [gr-qc]].

\bibitem{Mannheim:1991}
  P. D. Mannheim, D. Kazanas,
  ``Solutions to the Reissner-Nordstrom, Kerr, and Kerr-Newman problems in fourth-order conformal Weyl gravity,''
  Phys. Rev. D \textbf{44}, 417 (1991).

\bibitem{Villanueva:2013gga}
J.~R.~Villanueva and M.~Olivares,
``On the Null Trajectories in Conformal Weyl Gravity,''
JCAP \textbf{06} (2013), 040
[arXiv:1305.3922 [gr-qc]].

\bibitem{Turner:2020gxo}
G.~E.~Turner and K.~Horne,
``Null geodesics in conformal gravity,''
Class. Quant. Grav. \textbf{37} (2020) no.9, 095012

\bibitem{NewAMa}
  D. Z. Ma, X. Wu, J. F. Zhu,
  ``Velocity scaling method to correct individual Kepler energies,''
   NewA \textbf{13}, 216 (2008).

\bibitem{MaIJBC}
  D. Z. Ma, Z. C. Long, Y. Zhu,
  ``Application of indicators for chaos in chaotic circuit systems,''
   IJBC \textbf{26}, 11 (2016).

\bibitem{Benettin1976}
  G. Benettin, L. Galgani, J. M. Strelcyn,
  ``Kolmogorov entropy and numerical experiments,''
   Physical Review A, \textbf{14}, 2338 (1976).
   
\bibitem{Maldacena:2015waa}
J.~Maldacena, S.~H.~Shenker and D.~Stanford,
``A bound on chaos,''
JHEP \textbf{08}, 106 (2016)
[arXiv:1503.01409 [hep-th]].

\bibitem{Kitaev:2014}
A. Kitaev, ``Hidden correlations in the hawking radiation and thermal noise,'' talk given at the
Fundamental Physics Prize Symposium, Stanford University, Stanford, U.S.A., 10 November 2014.

\bibitem{Sekino:2008he}
Y.~Sekino and L.~Susskind,
``Fast Scramblers,''
JHEP \textbf{10}, 065 (2008)
[arXiv:0808.2096 [hep-th]].


\bibitem{Sachdev:1992}
S.~Sachdev, J. Ye, ``Gapless Spin-Fluid Ground State in a Random Quantum Heisenberg Magnet,'' Phys. Rev. Lett. \textbf{70}:3339,1993
[arXiv:cond-mat/9212030].

\bibitem{Kitaev:2015}
A. Kitaev, talks given at KITP, April and May 2015.

\bibitem{Sachdev:2015efa}
S.~Sachdev,
``Bekenstein-Hawking Entropy and Strange Metals,''
Phys. Rev. X \textbf{5}, no.4, 041025 (2015)
[arXiv:1506.05111 [hep-th]].

\bibitem{Stanford:2015owe}
D.~Stanford,
``Many-body chaos at weak coupling,''
JHEP \textbf{10}, 009 (2016)
[arXiv:1512.07687 [hep-th]].

\bibitem{Fu:2016yrv}
W.~Fu and S.~Sachdev,
``Numerical study of fermion and boson models with infinite-range random interactions,''
Phys. Rev. B \textbf{94}, no.3, 035135 (2016)
[arXiv:1603.05246 [cond-mat.str-el]].

\bibitem{Berenstein:2015yxu}
D.~Berenstein and A.~M.~Garcia-Garcia,
``Universal quantum constraints on the butterfly effect,''
[arXiv:1510.08870 [hep-th]].

\bibitem{Hashimoto:2016dfz}
K.~Hashimoto and N.~Tanahashi,
``Universality in Chaos of Particle Motion near Black Hole Horizon,''
Phys. Rev. D \textbf{95}, no.2, 024007 (2017)
[arXiv:1610.06070 [hep-th]].

\end{thebibliography}
\end{document}